\documentclass[acmsmall]{acmart}

\AtBeginDocument{%
  }

\acmBooktitle{Proceedings of the 32nd ACM Symposium on the Foundations of Software Engineering (FSE '24), July 15--19, 2024, Porto de Galinhas, Brazil}

\setcopyright{acmlicensed}
\acmDOI{10.1145/3643775}
\acmYear{2024}
\copyrightyear{2024}
\acmSubmissionID{fse24main-p1237-p}
\acmVolume{1}
\acmNumber{FSE}
\acmMonth{7}
\received{2023-09-29}
\received[accepted]{2024-01-23}

\usepackage{xcolor}

\usepackage{tabularx}
\usepackage{colortbl}

\usepackage{listings}
\usepackage{url}
\usepackage{hyperref}

\usepackage{graphicx}
\usepackage{array}			
\usepackage{multirow}
\usepackage{tabularx}
\usepackage{booktabs} 

\usepackage{enumitem}

\usepackage{tikz}

\usepackage{ifthen}
\newboolean{showcomments}
\setboolean{showcomments}{true} 
\ifthenelse{\boolean{showcomments}}
{\newcommand{\nb}[2]{
    \fcolorbox{gray}{yellow}{\bfseries\sffamily\scriptsize#1}
    {$\blacktriangleright$#2$\blacktriangleleft$}
  }
  
}
{\newcommand{\nb}[2]{}
  
}

\newcommand{\eg}{e.g.,~}							
\newcommand{\ie}{i.e.,~}							
\newcommand{\Fig}[1]{Figure~\ref{#1}}  			
\newcommand{\Table}[1]{Table~\ref{#1}}	    
\newcommand{\Sect}[1]{Section~\ref{#1}}	  
\newcommand{\oapp}{DISCOREV}

\usepackage{subcaption}

\begin{document}

\title{Improving the Learning of Code Review Successive Tasks with Cross-Task Knowledge Distillation}

\author{Oussama Ben Sghaier}
\orcid{0000-0003-2737-0952}
\affiliation{%
  \institution{Université de Montréal}
  \city{}
  \country{Canada}
}
\email{oussama.ben.sghaier@umontreal.ca}

\author{Houari Sahraoui}
\orcid{0000-0001-6304-9926}
\affiliation{%
  \institution{Université de Montréal}
  \city{}
  \country{Canada}
}
\email{sahraouh@iro.umontreal.ca}

\renewcommand{\shortauthors}{Oussama et al.}

\begin{abstract}

Code review is a fundamental process in software development that plays a pivotal role in ensuring code quality and reducing the likelihood of errors and bugs. However, code review can be complex, subjective, and time-consuming. \emph{Quality estimation}, \emph{comment generation}, and \emph{code refinement} constitute the three key tasks of this process, and their automation has traditionally been addressed separately in the literature using different approaches. In particular, recent efforts have focused on fine-tuning pre-trained language models to aid in code review tasks, with each task being considered in isolation. We believe that these tasks are interconnected, and their fine-tuning should consider this interconnection.
In this paper, we introduce a novel deep-learning architecture, named \oapp, which employs cross-task knowledge distillation to address these tasks simultaneously. 
In our approach, we utilize a cascade of models to enhance both \emph{comment generation} and \emph{code refinement} models. The fine-tuning of the \emph{comment generation} model is guided by the \emph{code refinement} model, while the fine-tuning of the \emph{code refinement} model is guided by the \emph{quality estimation} model. We implement this guidance using two strategies: a feedback-based learning objective and an embedding alignment objective. We evaluate \oapp~by comparing it to state-of-the-art methods based on independent training and fine-tuning. Our results show that our approach generates better review comments, as measured by the \emph{BLEU} score, as well as more accurate \emph{code refinement} according to the \emph{CodeBLEU} score.

\end{abstract}

\maketitle
\section{Introduction}

Code review is a pivotal process of the software development life cycle. Its primary purpose is to find issues, suboptimal code, and identify bugs \cite{mcintosh2014impact, mcintosh2016empirical}, all while ensuring the overall quality of the source code \cite{ackerman1989software, ackerman1984software, morales2015code}. This process primarily involves manual examination by one or more developers of code written by their peers \cite{fagan2002design, bavota2015four}. Code review encompasses several successive tasks, with the three most important being: \emph{quality estimation} of submitted code, description of potential issues (in the form of review comments), and \emph{code refinement} to address these issues.

\emph{Quality estimation} consists of assessing the quality of newly written code, with reviewers deciding whether to accept or reject a submitted pull request. Issue identification centers on pinpointing defects or problems within the source code, while review descriptions entail composing comments that not only highlight specific changes that introduced issues but also suggest potential solutions to address them \cite{mcintosh2014impact, mcintosh2016empirical}. \emph{Code refinement}, on the other hand, denotes the correction phase wherein developers address the identified issues to fix them, in accordance with reviewer comments \cite{bacchelli2013expectations}.

The code review process is often regarded as arduous, time-consuming, and intricate, especially in the context of large-scale projects \cite{eick2001does, avgeriou2016managing}. Moreover, it is a highly subjective process influenced by various human and social factors (\eg developers' experience, personal relationships, etc.). These factors can introduce biases in reviews, potentially leading to inefficiencies and inconsistencies in the code review process, ultimately impacting the quality and reliability of the codebase.

To address these challenges, there has been a growing interest in assistance and automation approaches to code review \cite{li2022automating, tufano2022using, tufan2021towards, siow2020core, gupta2018intelligent, hovemeyer2004finding}. One research area investigates the initial stages of the software development process by leveraging static analysis to detect potential issues \cite{wichmann1995industrial}. These tools, known as linters, define sets of manual rules representing various issues and flag sections of code that violate these predefined rules. However, the effectiveness of linters is limited due to their reliance on manually defined rules, which must be continually adapted to cover a range of issues. Additionally, software issues are subject to change over time and can be influenced by multiple factors such as software architecture, team culture, and project nature. Consequently, the rigid nature of static analysis tools limits their applicability and effectiveness in software projects \cite{bielik2017learning, sadowski2015tricorder}.

Other approaches have employed similarity techniques \cite{siow2020core, gupta2018intelligent} to recommend relevant comments from a pre-defined dataset of review comments based on similarities to code changes. While such suggestions may prove useful, review comments are rarely generic and more often context-specific.

With recent advances in deep learning and natural language processing, there has been significant interest in using pre-trained language models to address downstream tasks in software engineering. As a result, recent works \cite{tufano2022using, tufan2021towards, li2022automating, sghaier2023multi} have focused on using generative deep learning models to automate code review tasks, resulting in significant improvements in code review \emph{comment generation}. Despite the utility and promising results of these approaches, code review tasks have traditionally been considered separately, addressing them independently despite their significant interdependencies.

In this paper, we explicitly consider the interconnections among the three primary code review tasks: \emph{quality estimation}, \emph{comment generation}, and \emph{code refinement}. These tasks share knowledge, as the output of one task serves as input to another. The comment generated by the comment generation task feeds into the \emph{code refinement} task, while the revised code generated by the latter is used in the \emph{quality estimation} task to determine whether further review is necessary. In essence, \emph{comment generation} involves describing an issue to guide the subsequent \emph{code refinement} phase. \emph{Code refinement}, in turn, encompasses making the necessary code adjustments to align with the review comment and pass the \emph{quality estimation} phase.

Building upon this concept, we introduce a novel deep-learning architecture called \oapp~(\emph{cross-task knowledge DIStillation for COde REView}), which incorporates two models jointly trained on two related tasks. We apply this architecture to pairs of code review tasks, specifically \emph{comment generation/code refinement} and \emph{code refinement/quality estimation}.
Our proposed architecture is founded on cross-task knowledge distillation, where one model benefits from the feedback of another to achieve improved performance. \oapp~builds upon recent advancements in deep learning for code review, which have yielded promising results \cite{li2022automating, tufano2022using}. In addition to leveraging pre-trained models, our approach facilitates the joint modeling of code review tasks, enhancing overall effectiveness. The output of the second model informs the first model, providing feedback that gauges the relevance and informativeness of the generated output. Through the joint training of these two models, we can capture complex interactions, exchange knowledge and feedback, and ensure consistency between the two tasks. Furthermore, we introduce an embedding alignment objective to enforce closer semantic representations for the comment and code.

In evaluating \oapp, we conducted experiments using the same dataset of code reviews as employed in the literature \cite{li2022automating}. Our results demonstrate that our architecture outperforms state-of-the-art approaches for both \emph{comment generation} and \emph{code refinement} tasks, as measured by the \emph{BLEU} and \emph{CodeBLEU}scores. Additionally, we investigated the impact of the embedding alignment objective on model performance and found that enforcing closer semantic representations between comments and code edits enhances the \emph{comment generation} model's performance, resulting in more accurate and relevant comments.

The rest of this paper is structured as follows. 
\Sect{sec:background} provides a concise overview of the background.
\Sect{sec:approach} details the different components of our proposed approach.
\Sect{sec:evaluation} describes the evaluation results and discusses some threats to the validity of our results.
\Sect{sec:literature} discusses the related work.
Finally, \Sect{sec:conclusion} provides concluding remarks and outlines some ideas to improve our approach.

\section{Background}
\label{sec:background}

\subsection{Code review}
\label{Sec:code_review}
In a continuous integration setting, software developers utilize version control systems, such as GitHub, to collaborate and oversee software versions. 
As developers collaborate and persistently make changes to the codebase (\ie write new code or modify existing code), the use of version control systems enables development teams to efficiently manage and monitor the software codebase across time  \cite{shahin2017continuous, fowler2006continuous}.




In a continuous integration context, the conventional software development workflow is as follows.
A developer, who is working on implementing a new feature, makes local changes to the codebase.
Then, she commits and pushes her changes to a shared repository to make them available to collaborators.
She can create a pull request to ask that her changes be merged into the main branch. 
Some reviewers are assigned to the pull request to inspect these code changes. 
Reviewers identify and locate issues that need to be solved by the code author. 
The revised version of the code, submitted by developers, is reviewed again.
Once the changes are approved by reviewers, they are merged with the main code version to be available to other collaborators.
If necessary, the project can be deployed so that changes become available to end-users.

Code review is an essential task in the software development life cycle that aims to preserve the high quality of the software's codebase.
It entails regularly inspecting the source code written by fellow developers with the objective of identifying bugs, potential issues, sub-optimal code fragments, violations of code style, etc.
It is mainly composed of three tasks: \emph{quality estimation}, \emph{code review comments}, \emph{code refinement}.

\paragraph*{\textbf{Quality estimation}}
The nomenclature \emph{quality estimation} is adopted herein, in harmony with the work elucidated in \cite{li2022automating}, to maintain terminological consistency.
It is a pivotal step of the code review process that represents the assessment of a pull request, ultimately culminating in the consequential decision of either acceptance or rejection. 
During this process, the reviewer carefully checks the code to ensure it follows the rules, works correctly, and won't cause problems in the project. \emph{Quality estimation} helps maintain the overall quality of the software by only accepting changes that meet certain criteria.
If the pull request is accepted, the code changes are merged with the main branch of the codebase. Otherwise, the reviewer needs to explain the problems or the required improvements that the developer should consider. After making the necessary changes by the developer, the new code changes go through the process of \emph{quality estimation} again.

\paragraph*{\textbf{Code review comments}}
If the pull request is not accepted by the reviewer, she should describe the issues or improvements that need to be addressed.
Issues might be related to security vulnerabilities, code style, code standards violations, inconsistencies, bugs, sub-optimal code fragments, etc. Reviewers can also suggest improvements that involve code complexity, software libraries, best practices, refactoring, code quality, code optimization, documentation, etc. The identified issues and improvements are expressed by means of comments. The comments may be descriptive where the issues are detailed. However, comments may also be actionable where the reviewers recommend solutions to resolve the issues. 
The automation of this process helps to produce objective feedback without being influenced by personal preferences, relationships, health conditions, etc.

\paragraph*{\textbf{Code refinement}}
The comments that are produced by the previous step are fed as input to this phase. The developers, and authors of the code changes, should consider the feedback of the reviewers and perform the necessary code changes. Developers should solve the issues described in the comments and apply suggested solutions. New code changes are pushed again to the project repository and should satisfy the reviewer's comments. The refined version of the source code goes through the \emph{quality estimation} process again to check whether the comments are properly addressed. Proper \emph{code refinement} requires a good understanding of the reviewer's comments. That is, the reviews provided should be clear, specific, and comprehensible.

\subsection{Pre-trained language models}

In recent years, the Natural Language Processing (NLP) domain has dramatically evolved. 
State-of-the-art language models, that are based on the transformer architecture \cite{vaswani2017attention}, have shown outstanding performance in solving natural language problems.
Transformers have several advantages over recurrent neural networks. They enable parallel computations which reduces training time and captures long-range dependencies efficiently.
\emph{BERT}\cite{devlin2018bert}, \emph{GPT-3}\cite{radford2019language}, \emph{XLNet} \cite{yang2019xlnet}, \emph{RoBERTa} \cite{liu2019roberta} and \emph{T5} \cite{raffel2019exploring} are examples of general-purpose transformers that were pre-trained on tons of data to provide general and high-level representations of text.
\emph{CodeBERT} \cite{feng2020codebert} and \emph{CodeT5} \cite{wang2021codet5} are examples of transformers that were pre-trained on source code data.
These pre-trained models could then be adapted (\ie fine-tuned) on downstream tasks using specific datasets (\eg fine-tune a language model to predict if a code fragment contains bugs).

Transformers are sequence-to-sequence models based on the attention mechanism that takes into account the relationship between all the words in the sentence and not only the contextual words (\eg previous and next words).
The attention mechanism learns a weighting function indicating how much each element contributes to the prediction (\ie importance of each input element in predicting the target). 
A transformer is composed of encoders and decoders.
The encoder transforms the input sequence into contextual embeddings. 
These latter encode the meaning of each word of the input based on its context in low-dimension and high-level representations.
The decoder uses these contextual representations to predict the output sequence.

\emph{CodeT5} \cite{wang2021codet5} is a pre-trained encoder-decoder transformer for code that is pre-trained on $8.35M$ functions in 8 programming languages.
There are several versions of \emph{codeT5}. The small version has $60$ million parameters and the base version has $220$ million parameters.

\subsection{Knowledge distillation}
\label{Sec:distillation}
Knowledge distillation is a popular technique that was first introduced by Hinton et al. \cite{hinton2015distilling} as a method to transfer knowledge from a complex model, also known as the teacher, to a smaller and faster model, called the student. The main goal of knowledge distillation is to transfer the learned knowledge of the teacher model to the student model so that it can achieve comparable or even better performance on a target task.

The process of knowledge distillation consists of training a teacher model on a large dataset to accomplish a specific task. Then, the student model is trained to mimic the behavior of the teacher model by minimizing the distance between their output distributions on the same dataset, typically using the Kullback-Leibler divergence (KL-divergence) as the distance metric \cite{joyce2011kullback, kim2021comparing}.

For instance, this concept is employed in model compression where the knowledge is transferred from a large and complex neural network (\ie the teacher) to a smaller and simpler neural network (\ie the student). The goal of knowledge distillation is to make the student network learn the same function as the teacher network but with fewer parameters. 

Cross-task knowledge distillation is a recent extension of knowledge distillation that enables the transfer of knowledge from one task to another\cite{ye2020distilling, yuan2019ckd, yang2022cross, li2022prototype}. Instead of transferring knowledge from a complex model to a smaller one for the same task, cross-task knowledge distillation transfers knowledge from a model trained on one task, referred to as the source task, to a model trained on a different task, referred to as the target task. This allows the target model to improve its performance, even if the two tasks are not directly related.

\section{Proposed approach}
\label{sec:approach}

In this section, we introduce our proposed architecture \oapp~in a broad and abstract context, elucidating its relevance for software engineering tasks. 
Subsequently, we will delve into its specific instantiation for code review tasks.

\subsection{A learning architecture using cross-task knowledge distillation}

Most deep learning models are trained using their own feedback, which involves comparing their predictions to the true labels. Nevertheless, some approaches utilize feedback from another neural network to enhance training. In this way, the training signal becomes more informative than a simple comparison with the ground truth.

Jointly training two models can take the form of competition or collaboration. Competitive models compete to outperform each other based on certain metrics. Conversely, collaborative models cooperate through a feedback mechanism to achieve better performance. For example, generative adversarial networks (GANs) are employed for generative tasks where the objective is to create new data resembling the original data distribution \cite{goodfellow2020generative}. The typical GAN architecture consists of two neural networks: a generator and a discriminator. The generator learns to create synthetic data, while the discriminator learns to distinguish between the generated and real data. GANs are trained in an adversarial and competitive manner, with the generator attempting to fool the discriminator, while the discriminator strives to improve its ability to differentiate between real and fake data. Consequently, both models improve with training over time. In contrast, knowledge distillation is a collaborative learning technique that involves transferring knowledge from a larger model to a smaller one, as elucidated in \Sect{Sec:distillation}. 

In software engineering, tasks are traditionally addressed independently. Thus, to handle each task, a deep learning model is either fine-tuned or trained from scratch. However, many software engineering tasks are interconnected, as the output of one task may serve as the input for another. Instead of approaching these tasks in isolation, we propose a paradigm shift that capitalizes on the inherent interplay between them.
The essence of our proposition lies in the relationship between sequentially executed software engineering tasks, which can be leveraged as a potent force for enhanced task automation. By orchestrating these tasks harmoniously within a shared framework, we aim to harness common knowledge and feedback between tasks. We conjecture that this approach will lead to more robust, coherent, and efficient trained models that inherently capture the dependencies and intricacies among the constituent tasks.

\begin{figure*}[!htbp]
    \centering
    \includegraphics[width=0.85\textwidth]{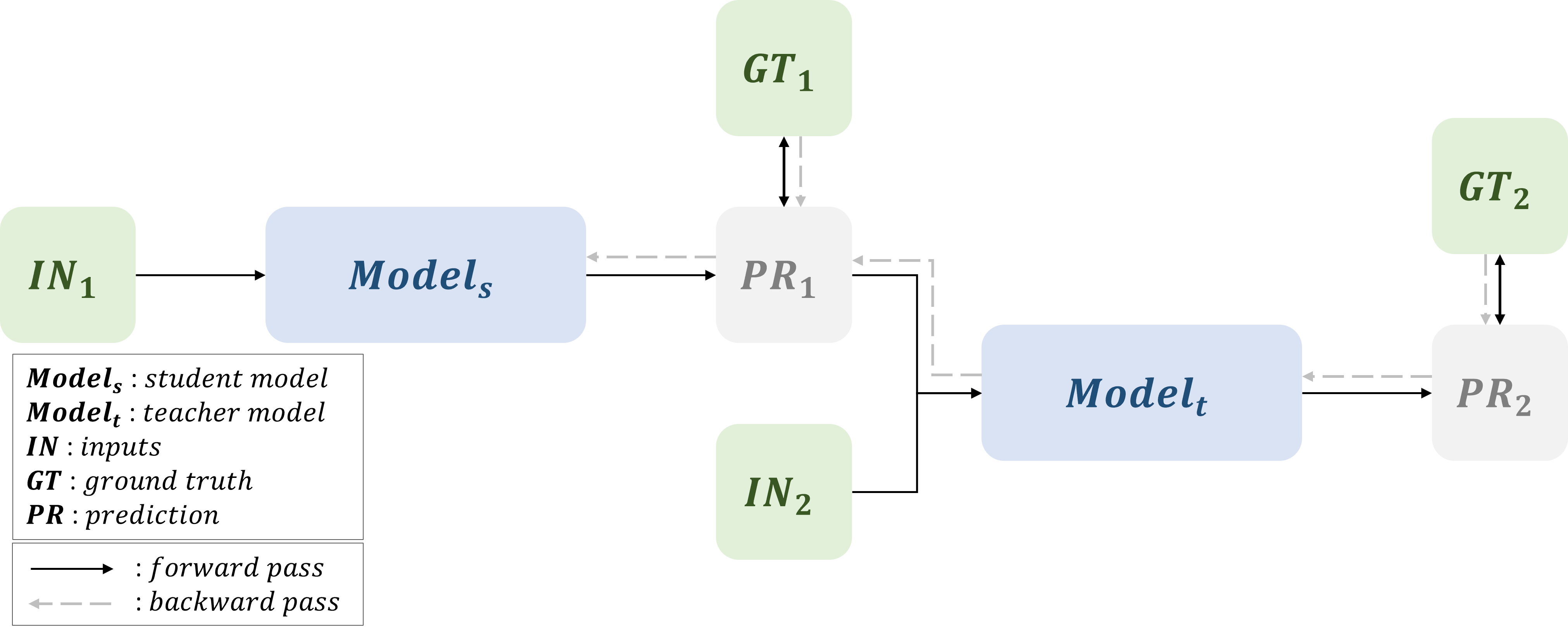}
    \caption{Overview of the proposed architecture \oapp.}
    \label{fig:abstract-architecture}
\end{figure*}

We introduce \oapp, a novel deep learning architecture based on cross-task knowledge distillation. \oapp~is designed to address distinct but interconnected tasks, particularly within the software development domain. Our effort involves the cohesive modeling of these tasks, as abstractly illustrated in \Fig{fig:abstract-architecture}. 
Within this architecture, we have conceived two interconnected models that operate symbiotically, informed by feedback mechanisms to accomplish both tasks. Through joint training, our architectural design facilitates an efficient learning process, allowing the first model to be guided by the second model. 

More specifically, let's consider the scenario where we are dealing with two interdependent tasks $T_1$ and $T_2$, where the output of $T_1$ serves as the input to $T_2$. In this context, $Model_s$ is the student model responsible for automating $T_1$, while $Model_t$ is the teacher model that addresses $T2$.
The underlying intuition is that, in addition to comparing predictions to the ground truth, the model for the subsequent task provides guidance to the first model through feedback. This feedback serves as an informative signal emanating from the second model, offering the first model insight into the relevance of its output and its contribution to fulfilling the second task.

$Model_s$ takes some inputs $IN_1$ and generates a prediction $PR_1$ which is then passed, along with other potential inputs $IN_2$ to the second model $Model_t$.
The teacher model subsequently generates its own prediction $PR_2$.
The loss function of $Model_t$ is defined based on the comparison between its prediction $PR_2$ and the ground truth $GT_2$.
The loss function of the student model $Model_s$ consists of two components: a regular loss function that compares its prediction $PR_1$ to the ground truth $GT_1$, and the loss of the teacher model. Consequently, the optimization objective for the student model extends beyond the singular minimization of its own loss. It takes into account the loss incurred by the teacher model, aiming to generate predictions that not only approximate the ground truth but also facilitate the teacher model in achieving precision in its predictions. Therefore, during backpropagation, the weights of the student model are adjusted to optimize both tasks simultaneously.

As mentioned earlier, in the context of code review, tasks have traditionally been treated separately in the literature \cite{li2022automating, tufano2022using, tufan2021towards}, with a predominant focus on the local scope, \ie comparing the generated output to the true labels, despite their clear and inherent interdependence. In our work, we postulate that code review tasks are closely intertwined, as producing accurate code edits should depend on the output reviews, and estimating the code quality relies on the generated code edits. Therefore, we apply \oapp~to enhance the performance of deep learning models in automating code review tasks. In the following sections, we illustrate this application by following the chain of dependency from teacher to student. That is, we begin with \emph{quality estimation} guiding \emph{code refinement}, and then proceed with \emph{code refinement} guiding \emph{comment generation}.

\subsection{Code refinement guided by quality estimation using \oapp}

\begin{figure}[!htbp]
\begin{subfigure}{1\textwidth}
  \centering
  \includegraphics[width=0.9\linewidth]{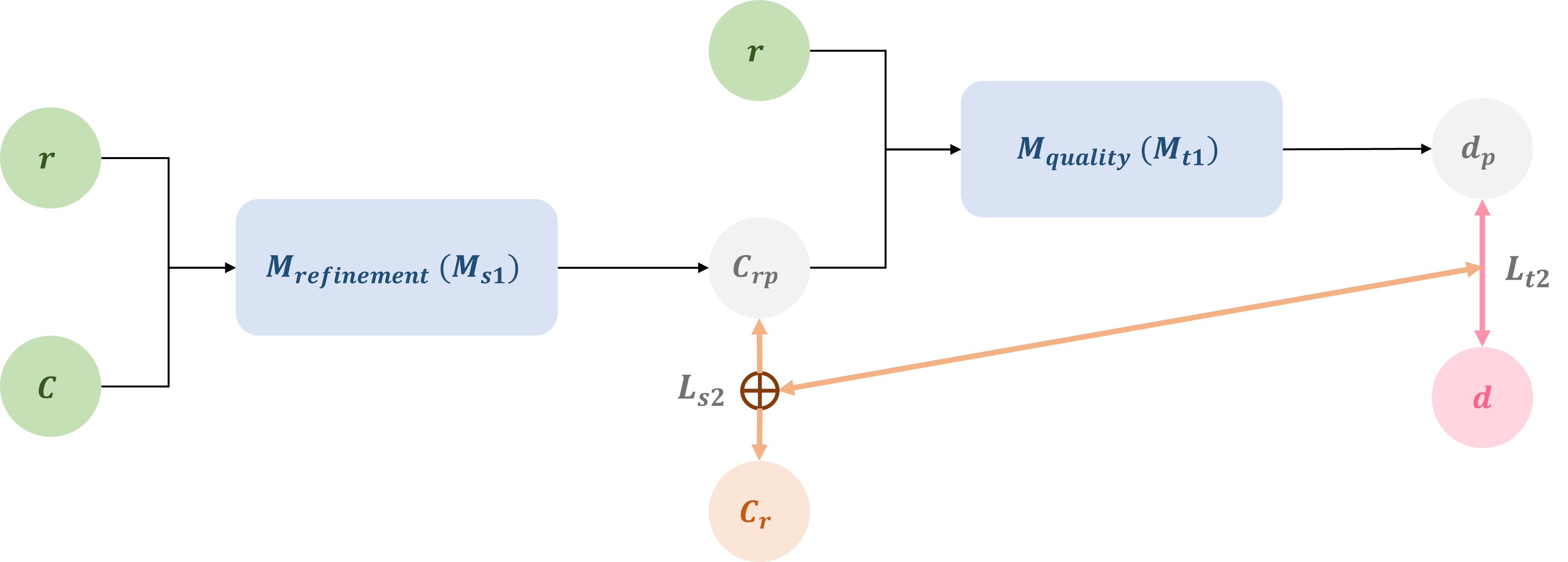}
  \caption{\oapp~applied to the code refinement and quality estimation tasks}
  \label{fig:architecture_details2}
\end{subfigure}
\\ \vspace{10pt}
\begin{subfigure}{1\textwidth}
  \centering
  \includegraphics[width=0.9\linewidth]{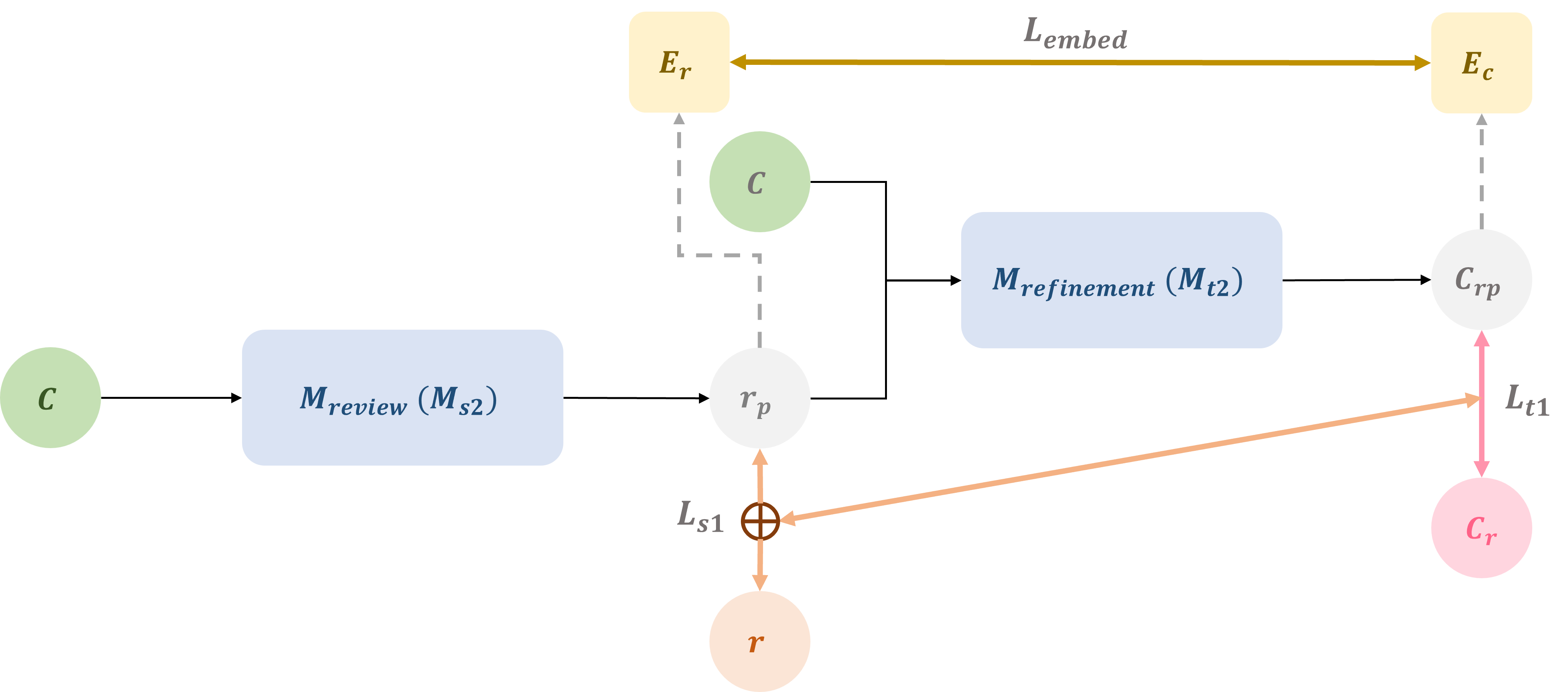}
  \caption{\oapp~applied to comment generation and code refinement tasks}
  \label{fig:architecture_details1}
\end{subfigure}
\caption{Detailed architecture of \oapp~in the context  of code review.}
\label{fig:architecture}
\end{figure}

In this part, we rely on the interdependence between the \emph{code refinement} and \emph{quality estimation} tasks.
During the \emph{code refinement} task, a developer tries to make appropriate changes to the code base with the aim of getting the pull request accepted in the next task, \ie \emph{quality estimation}. 
As shown in \Fig{fig:architecture_details2}, we instantiated \oapp~to address these tasks jointly. 

We denote $\mathcal{D}_1$ our dataset that is composed of pairs $(c, d)$, where $c$ is a code snippet and $d$ is a binary value that indicates whether the code should be accepted or rejected.
Given a review comment $r$ and a code snippet $c$ from $\mathcal{D}_1$, the student model $\mathcal{M}_{s1}$ generates the code edits $c_{rp}$ (\ie refined version of the code), an approximation of $c_r$. 
The generated code edits $c_{rp}$ are fed, along with the comment, to the teacher model $\mathcal{M}_{t1}$ that outputs the probability $d_p$ of the code changes to be accepted or rejected. That is, this value indicates the extent to which the code edits satisfy the review comment given as input.
The loss gradients of $\mathcal{M}_{t1}$ backpropagates to $\mathcal{M}_{s1}$. We consider this as feedback from the teacher model that tells the relevance and conformance of the code edits to the review comment.

$\mathcal{M}_{t1}$ and $\mathcal{M}_{s1}$ are defined, respectively, as follows: 
\begin{equation}
\left\{
    \begin{array}{cc}
        \mathcal{M}_{s1}:  & c, r \rightarrow c_{rp} \\
        \mathcal{M}_{t1}: & c_{rp}, r \rightarrow d
    \end{array}
\right.
\end{equation}

The loss of the teacher model $\mathcal{L}_{t1}$ is defined by:
\begin{equation}
    \mathcal{L}_{t1} = \mathcal{L}_{BCE}(d, d_p)
\end{equation}
It consists of comparing the truth value $d$ with the predicted value of $\mathcal{M}_{t1}$ using the binary cross-entropy loss function $\mathcal{L}_{BCE}$ that is defined with the following equation:
\begin{equation}
        \mathcal{L}_{BCE}(y, \hat{y}) = y \log(\hat{y}) + (1-y) \log(1-\hat{y})
\end{equation}
where $\mathcal{L}_{BCE}$ represents the binary cross-entropy loss, $y \in \{ 0,1 \}$ is the truth value, and $\hat{y}$ is the probability predicted by the classifier.


The loss of the student model $\mathcal{L}_{s2}$ is defined as a combination of the cross-entropy loss and the teacher model loss.
\begin{equation}
    \mathcal{L}_{s1} = \alpha \mathcal{L}_{CE}(\mathcal{P}_{c_r}, \mathcal{P}_{c_{rp}}) + \beta \mathcal{L}_{t1}
\end{equation}
where $\alpha$ and $\beta$ are weights that indicate the importance of each loss.

The cross-entropy loss compares the probability distributions of the predicted code edits $\mathcal{P}_{c_r}$ and the real ones $\mathcal{P}_{c_{rp}}$.
The formula that defines the cross-entropy loss is given by:
\begin{equation}
    \mathcal{L}_{CE}(p, q) = -\sum_{i=1}^{N} p_i \log(q_i)
\end{equation}
where $\mathcal{L}_{CE}$ represents the cross-entropy loss, $N$ is the number of tokens in the sentence, $p_i$ is the one-hot encoded vector representing the true $i^{th}$ token, $q_i$ is the predicted probability vector of the $i^{th}$ token. The vector is of size $|vocabulary|$ and indicates for each word in the vocabulary its probability to be the $i^{th}$ token.


The teacher model provides feedback to the student model to indicate whether the generated code edits conform to the review comment (\ie accept or reject the pull request).
That is, the student model will try not to only generate code edits that are close to the ground truth but have also a higher chance of being accepted by the second model.

\subsection{Comment generation guided by code refinement using \oapp}

In code review, \emph{comment generation} and \emph{code refinement} tasks are strongly related.
The reviewer's comment guides the \emph{code refinement} task and describes what the developers should do to improve/fix their code.
On the other hand, developers refine their code by executing the instructions of the review trying to satisfy the reviewer's comments.
We apply \oapp~ on these two tasks to address them simultaneously.

\Fig{fig:architecture_details1} illustrates the details of our unified architecture for the \emph{comment generation} and \emph{code refinement} tasks.
The student model generates a review comment for the source code given as input.
The generated review is given, along with the code, as input to the teacher model to generate the necessary code edits that fix the input code with respect to the review. 
The loss gradients of the teacher model backpropagate to the first model. That is, the teacher model gives feedback to the student model to indicate the relevancy of the generated review and to what extent it enables the teacher model to generate the appropriate code edits.

Formally, we denote $\mathcal{D}_2$ our dataset that is composed of triplets $(c, r, c_r)$ where $c$ is the initial version of the source code submitted by the developer for review, $r$ is the review comment, and $c_r$ is a refined version of $c$ with respect to $r$.
$\mathcal{M}_{t2}$ and $\mathcal{M}_{s2}$ refer to the teacher and student models, respectively, such that: 
\begin{equation}
\left\{
    \begin{array}{cc}
        \mathcal{M}_{s2}:  & c \rightarrow r_p \\
        \mathcal{M}_{t2}: & c, r_p \rightarrow c_{rp}
    \end{array}
\right.
\end{equation}

The student model takes as input $c$ and generates the review $r_p$.
$r_p$ is fed along with $c$ to the teacher model that generates $c_{rp}$ trying to refine the source code.
The loss of the teacher model $\mathcal{L}_{t2}$ is defined as follows:
\begin{equation}
    \mathcal{L}_{t2} = \mathcal{L}_{CE}(\mathcal{P}_{c_r}, \mathcal{P}_{c_{rp}})
\end{equation}
It consists of comparing the predicted code edits to the real ones using the cross-entropy loss function $\mathcal{L}_{CE}$ that compares the probability distribution $\mathcal{P}$ of the predicted output to the ground truth.

The loss of the student model $\mathcal{L}_{s2}$ is defined as follows:
\begin{equation}
    \mathcal{L}_{s2} = \alpha \mathcal{L}_{CE}(\mathcal{P}_{r}, \mathcal{P}_{r_{p}}) + \beta \mathcal{L}_{t2}
\end{equation}
where $\alpha$ and $\beta$ indicate the importance of each loss.

It is a combination of the common cross-entropy loss and the loss of the teacher model.
In short, the loss of $\mathcal{M}_{s2}$ compares the predicted review to the real one, but also considers the feedback of the teacher.
That is, the student model will try not only to generate reviews that are close to the ground truth but also relevant reviews that enable the second model to predict the right code edits.



\paragraph{\textbf{Embeddings alignment objective}}

Until now, we have been considering the loss of the \emph{code refinement} model within one of the \emph{comment generation} model. However, we can capitalize on another property of our problem to enhance interconnection-based learning. In our approach, the models responsible for generating reviews and code edits are closely intertwined, given that reviews often describe the necessary code edits, and conversely, code edits typically represent the required alterations to the code to align with the review comment. Therefore, a review comment and its corresponding code edits inherently share similar meanings. 
This close relationship suggests the existence of shared underlying patterns or features, and consequently, similar representations, within the embeddings of reviews and code edits.

In natural language processing, hidden representations refer to intermediate representations of textual data learned by deep learning models during training.
These representations are not directly interpretable by humans, as they usually consist of high-dimensional vectors capturing intricate patterns and relationships among words and phrases in the input text. Within the context of language models, each hidden layer may encode different aspects of the input text, such as syntactic structure, semantic meaning, or discourse coherence \cite{lopez2022ast}.
The encoding of each hidden layer can vary in several ways, including the level of abstraction, granularity of the representations, amount of contextual information captured, or specialization for specific tasks or domains. For instance, lower-level layers in a language model may encode fundamental linguistic features like word-level embeddings or syntactic parse trees, while higher-level layers may capture more abstract semantic or discourse-level connections between words and sentences. 
Typically, the last hidden layer of the encoder serves as an embedding for the input text, encapsulating its meaning. This embedding is utilized by the decoder to generate the desired output. Developing models capable of harnessing effective representations can have implications for downstream tasks such as text classification or language generation.

To enhance the performance of our models, we introduce an \emph{embeddings alignment objective} aimed at aligning the embeddings of the review comment and the code edits. The objective is to achieve close representations, as review comments and code edits share the same meaning. \Fig{fig:architecture_details1} illustrates this objective.
We use codeT5 for models $M_{s2}$ and $M_{t2}$, both of which are based on the (encoder-decoder) transformer architecture. 
The representations generated by the encoder of $M_{s2}$ are employed as embeddings for the predicted review, while the embeddings generated by the encoder of $M_{t2}$ are utilized for the predicted revised version of the code.

Let us assume a predicted review $r_p$ and its corresponding predicted code edits $c_{rp}$ whose vector representations are $E_r$ and $E_c$ of size $n$. To align these embeddings, we define an objective $\mathcal{L}_{embed}$ that minimizes the distance between the two vectors defined as their mean square error (MSE):
\begin{equation}
\mathcal{L}_{embed} = MSE(E_r, E_c) =  \frac{1}{n} \sum_{i=1}^n \left({E_c}_i - {E_r}_i\right)^2
\end{equation}

Our objective is to minimize the embedding alignment loss $\mathcal{L}_{embed}$ to achieve better and shared representations.
Thus, we incorporate it in the teacher and student losses as follows:
\begin{equation}
\mathcal{L'}_{t_2} = \alpha_1 \mathcal{L}_{t2} + \beta_1 \mathcal{L}_{embed}
\end{equation}
\begin{equation}
    \mathcal{L'}_{s2} = \alpha_2 \mathcal{L}_{s2} + \beta_2 \mathcal{L}_{embed}
\end{equation}
where $\alpha_i$ and $\beta_i$ are weights that denote the importance of each loss. 
For the new student loss~$\mathcal{L'}_{s2}$,  the teacher loss is incorporated in $\mathcal{L}_{s2}$.

\section{Evaluation}
\label{sec:evaluation}

We conducted a series of experiments to assess the performance of our proposed architecture\footnote{The replicate package is provided in the supplementary material of this paper.}. In this section, we first outline the research questions we sought to address, followed by a description of the experimental setup. Subsequently, we present the results and discuss various potential threats associated with our experiments.

\subsection{Research questions}

To assess \oapp, we formulate six research questions. The first question pertains to the training of the \emph{code refinement} model, which is guided by the \emph{quality estimation} model, the \emph{quality estimation} model being a straightforward binary classifier. The remaining five questions focus on training the \emph{comment generation} model, which is guided by the \emph{code refinement} model. In this case, the teacher model is more complex, and the relationship between the student and the teacher necessitates a deeper analysis.

\begin{itemize}
    \item \textbf{RQ1:} \textbf{Accuracy on code refinement}\\
        \emph{How does the \emph{code refinement} model, guided by the \emph{quality estimation} model, perform compared to baseline models?}
    
    \item \textbf{RQ2:} \textbf{Accuracy on comment generation}\\
        \emph{How does the \emph{comment generation} model, guided by the \emph{code refinement} model (and the embedding alignment), perform compared to baseline models?}

    \item \textbf{RQ3:} \textbf{Effect of the embedding alignment objective}\\
        \emph{Considering that the inclusion of the embedding has an effect on the training cost, do we have similar results without considering the embedding?}

    \item \textbf{RQ4:} \textbf{Training time of \oapp}\\
        \emph{What is the additional time of the joint training compared to baseline models?}
\end{itemize}
Taking into account the training time, we considered two additional questions for the simple version of \oapp.
\begin{itemize}    
    \item \textbf{RQ5:} \textbf{Accuracy on comment generation for each programming language}\\
        \emph{How does the same model perform, compared to baseline models, on \emph{comment generation} with respect to each programming language?}
    
    \item \textbf{RQ6:} \textbf{Impact of the teacher pre-finetuning}\\
        \emph{To what extent does the pre-finetuning phase of the teacher contribute to enhancing the accuracy of the student model?}

\end{itemize}


\subsection{Experimental setup}

\paragraph{\textbf{Dataset}}
We use a dataset of $176~616$ code reviews obtained from \cite{li2022automating}. 
The dataset has code review entries for nine distinct programming languages.
\Table{tab:dataset} provides an overview of the dataset.
The data is split into $85\%$ for \emph{Train}, $7.5\%$ for \emph{Validation}, and $7.5\%$ for \emph{Test}. 
The data contains more than $1.3M$ lines of code.
To have a fair and consistent comparison with state-of-the-art works, we use the same data split as in \cite{li2022automating}. 
We cleaned the dataset and we kept the necessary features as explained below.
The first resulting dataset is a set of triplets $(c, r, c_r)$ where $c$ is the initial version of the code, $r$ is the review comment, and $c_r$ is the revised source code after addressing the comment. The second dataset is a set of triplets $(c_c, r, d)$ where $d$ is a binary value indicating whether the candidate code changes $c_c$ should be accepted or rejected according to comment $r$.


\begin{table}[!htbp]
  \centering
  \caption{Distribution of the dataset over programming languages}
  \label{tab:dataset}
  \begin{tabular}{lccccccccc|c}
    \toprule
    & \textbf{PHP} & \textbf{Ruby} & \textbf{C\#} & \textbf{C} & \textbf{Java} & \textbf{Python} & \textbf{C++} & \textbf{Go} & \textbf{JS} & \textbf{Total} \\
    \midrule
    \textbf{Train} & 7 979 & 5 730 & 15 630 & 3 077 & 31 288 & 30 648 & 13 274 & 30 369 & 12 411 & 150 409 \\
    \textbf{Val} & 973 & 504 & 717 & 543 & 2 177 & 2 834 & 1 362 & 2 865 & 1 128 & 13 103 \\
    \textbf{Test} & 1 032 & 479 & 738 & 488 & 2 206 & 2 900 & 1 308 & 2 889 & 1 064 & 13 104 \\
    \bottomrule
  \end{tabular}
\end{table}

\newcommand{\bos}{~<\!s\!>~}
\newcommand{\eos}{~<\!/s\!>~}
\newcommand{\msg}{~<\!msg\!>~}
\newcommand{\pad}{~<\!pad\!>~}

\paragraph{\textbf{Preprocessing}}
To properly train \oapp, the various inputs and outputs should be represented conveniently.

The \emph{comment generation} model is fed with the code $c$ as input and generates a review $r_p$.
We clean the input code $c$ and use RoBERTa tokenizer \cite{liu2019roberta} to split it into tokens.
\[
 c \Rightarrow c_1 c_2 \dots c_n
\]
The inputs are either truncated or padded with a special token $\pad$ to assert a fixed input length.
We add two other special tokens $\bos$ and $\eos$ to denote the beginning and the end of the input respectively.
The resulting input has this representation:
\[ 
c_1 c_2 \dots c_n \Rightarrow  \bos c_1 c_2 \dots c_n\pad\dots\pad~\eos
\]
where $c_i$ is the $i^{th}$ code token.

The output $r_p$ is also preprocessed in the same way:
\[ 
r \Rightarrow \bos r_1 r_2 \dots r_n\pad..\pad~\eos
\] 
where $r_i$ is the $i^{th}$ review token.

The \emph{code refinement} model takes as input the predicted review along with the code and generates the revised version of the code.
The inputs are also cleaned, truncated or padded, and augmented with the special tokens.
The resulting input is represented as follows: 
\[ 
(c, r) \Rightarrow \bos c_1 c_2 \dots c_n\pad\dots\pad~\msg r_1 r_2 \dots r_m \pad\dots\pad~\eos
\] 
where $\msg$ is a special token that separates the input code and review.
The output is preprocessed similarly as follows: 
\[ 
c_r \Rightarrow \bos c_{r_1} c_{r_2} \dots c_{r_n}\pad..\pad~\eos
\]

\paragraph{\textbf{Experiments}}

In \emph{RQ1}, we assessed our approach's performance on \emph{code refinement}. Initially, we pre-finetuned the \emph{quality estimation} model to adapt it to the downstream task. Subsequently, we performed joint training of \oapp~on the \emph{code refinement} and \emph{quality estimation} tasks, as depicted in \Fig{fig:architecture_details2}. During this phase, the student model (\ie \emph{code refinement} model $\mathcal{M}_{s1}$) underwent fine-tuning using its own loss function $\mathcal{L}_{s1}$ while receiving feedback from $\mathcal{M}_{t1}$ through the loss $\mathcal{L}_{t1}$.
For $\mathcal{M}_{t1}$, we compared its prediction with $1$ to enforce the student model to generate code edits that would be accepted by the teacher model.
To prevent the teacher model from learning to consistently predict $1$ simply to minimize its loss, we kept its weights frozen. Consequently, the student model is responsible for minimizing both losses (\ie generating code edits that align with the expected changes and ensuring the pull request was accepted). We assessed the performance of the resulting \emph{code refinement} model on the test set and compared it to the baseline. 

For \emph{RQ2}, we similarly fine-tuned \oapp~on the \emph{code refinement} and \emph{comment generation} tasks.
In the first phase, we fine-tuned $\mathcal{M}_{t2}$ on \emph{code refinement} to equip the teacher model with prior knowledge for providing informative and accurate feedback to the student model. In the subsequent phase, we jointly fine-tuned models $\mathcal{M}_{s2}$ and $\mathcal{M}_{t2}$, as depicted in \Fig{fig:architecture_details1}. For each model, we extracted the output of the last hidden layer of the encoder as embeddings.
We evaluated our generated models on the test set and compared their performance to that of state-of-the-art works. 

For \emph{RQ3}, to assess the impact of the embeddings alignment objective ($\mathcal{L}_{embed}$), we omitted this component and re-ran the previous experiment.

To answer \emph{RQ4}, we ensured that all models were trained using the same resources (\ie GPUs and memory). 
We recorded the time required for each model to complete one epoch.

For a deeper investigation into our proposed architecture, we selected the \emph{comment generation} and \emph{code refinement tasks} to conduct additional experiments.
In \emph{RQ5}, we evaluated the performance of \oapp~across different programming languages by using the separate datasets shown in \Table{tab:dataset}.

In \emph{RQ6}, we explored the impact of the pre-finetuning phase through two experiments. First, we pre-finetuned the teacher model on the \emph{code refinement} task before fine-tuning \oapp. Second, we fine-tuned the \oapp~model without pre-finetuning the teacher model. We evaluated the resulting models on the test set and compared the two alternatives.

\paragraph{\textbf{Baseline models}}
We compare \oapp~with state-of-the-art works: T5 \cite{tufan2021towards}, CodeT5 \cite{wang2021codet5}, and CodeReviewer \cite{li2022automating}.
In \cite{tufan2021towards}, the authors fine-tuned T5 on \emph{comment generation} and \emph{code refinement}.
We denote CodeT5 as the fine-tuned version of this model for downstream tasks.
In \cite{li2022automating}, the authors introduced \emph{CodeReviewer}, a codeT5 model that was pre-trained on some tasks related to code review, and then fine-tuned on \emph{comment generation}, \emph{code refinement}, and \emph{quality estimation} tasks.

\paragraph{\textbf{Performance metrics}}
To assess the effectiveness of the models, we compute the Bilingual Evaluation Understudy (BLEU) score \cite{papineni2002bleu} for the generated comments and CodeBLEU score \cite{ren2020codebleu} for the generated code.
The BLEU score is a widely used metric for evaluating the quality of text generated by deep learning models. 
It measures the degree of similarity between the generated text and a set of reference translations, in our case, the generated review and the actual review. A higher BLEU score indicates better text quality with respect to the references.
We use the BLEU-4 variant, that computes the n-gram overlap ($1\leq n\leq 4$) using this formula:
\begin{equation*}
BLEU\!-\!4 = min(1, \frac{output\_length}{reference\_length})(\prod_{i=1}^{4}precision_i)^{\frac{1}{4}}    
\end{equation*}
It computes the precision for n-grams of size 1 to 4 and adds a brief penalty for short sentences. 
CodeBLEU is an adaptation of BLEU tailored for code.
It combines the n-gram match (computed by the BLEU score), weighted n-gram match, syntactic AST match, and semantic data flow match.
These metrics are also used in \cite{tufano2022using, li2022automating} which sets a consistent and fair basis for comparison.

\paragraph{\textbf{Implementation and parameters}}
To implement the proposed models and run the different experiments, we used the PyTorch framework. 
We run the training on a machine with four GPUs \emph{RTX 3090} and $24 GB$ of RAM per GPU. The models were trained using the Adam optimizer with a learning rate of $10^{-5}$, and we used cross-entropy as a loss function. 
The training process was performed on the training set for $30$ epochs, with a batch size of $16$. The validation set was used to monitor the model's performance and adjust the different hyperparameters accordingly.
Lastly, the test set was used in the final step to evaluate the performance of the models. 
We equally set the parameters $\alpha$ and $\beta$ to $\frac{1}{2}$.

\subsection{Results}


\subsection*{Results for RQ1 - Accuracy on code refinement}

\Table{tab:results5} shows the results of this research question.
In a first phase, we fine-tuned our model on the task of code change \emph{quality estimation} to accurately predict if a pull request should be accepted or rejected given the input code changes. Within this preliminary phase, a good accuracy of $71.46$ was attained, which sets a good knowledge base for the \emph{quality estimation} model to guide the \emph{code refinement} model.

Then, we fine-tuned \oapp~on both tasks simultaneously.
As shown in \Table{tab:results5}, \oapp~outperforms the baseline models on \emph{code refinement}.
It reaches $85.49$ as CodeBLEU score compared to $82.61$ for CodeReviewer, the highest score of the baselines. 
This confirms our earlier conjecture, suggesting that simultaneously addressing code review tasks could improve the accuracy of the learning. 


\begin{table}[!htbp]
  \centering
  \caption{Performance of \oapp~ on code refinement compared to baseline models}
  \label{tab:results5}
    \begin{tabular}{l c c}
    \toprule
    \textbf{Model} & \textbf{Code refinement} & \textbf{Quality estimation}\\
    \midrule
    \rowcolor{black!10} \multicolumn{3}{c}{\textcolor{black}{\textbf{Pre-finetuning}}}\\
        \oapp & & 71.46\\
    \rowcolor{black!10} \multicolumn{3}{c}{\textcolor{black}{\textbf{Fine-tuning}}}\\
        T5 & 77.03 & \\
        CodeT5 & 80.82 & \\
        CodeReviewer & 82.61 & 73.89\\
        \oapp & \textbf{85.49} & 71.46\\
    \bottomrule
\end{tabular}
\end{table}


\subsection*{Results for RQ2 - Accuracy on comment generation}

\Table{tab:results1} reports the results obtained for the \emph{comment generation} task.
In the pre-finetuning, the \emph{code refinement} model achieved a codeBLEU score of $81.79$.
The produced model was used in the fine-tuning phase of \oapp~to guide, through feedback, the student model on the \emph{comment generation} task.
As illustrated in \Table{tab:results1}, \oapp~outperforms the baseline models in the \emph{comment generation} task, registering remarkable improvements in BLEU scores of $67\%$, $52\%$, and $38\%$ over the T5, CodeT5, and CodeReviewer baselines, respectively.

The discernible difference in accuracy for \emph{comment generation} shows the effectiveness of our proposed feedback-based learning strategy when the pre-trained \emph{code refinement} model provides valuable guidance for the \emph{comment generation} task.


In our approach, we do not anticipate an enhancement in the accuracy of the teacher model through joint training. This is primarily because the \emph{code refinement} model is fine-tuned with synthetic reviews generated by the \emph{comment generation} model, as opposed to actual reviews. These generated reviews may potentially contain grammatical errors and lack precision. Furthermore, the feedback process is unidirectional, meaning that the teacher model does not receive direct feedback from the student model. Nevertheless, it is noteworthy that the \emph{code refinement} model demonstrated a marginal improvement in accuracy. We conjecture that this improvement is attributed to the embedding alignment objective.


\begin{table}[!htbp]
  \centering
  \caption{Performance of \oapp~on comment generation compared to baseline models}
  \label{tab:results1}
  \begin{tabular}{l c c}
      \toprule
    \textbf{Model} & \textbf{Comment generation} & \textbf{Code refinement}\\
    \midrule
    \rowcolor{black!10} \multicolumn{3}{c}{\textcolor{black}{\textbf{Pre-finetuning}}}\\
        \oapp & & 81.79\\
    \rowcolor{black!10} \multicolumn{3}{c}{\textcolor{black}{\textbf{Fine-tuning}}}\\
        T5 & 4.39 & 77.03\\
        CodeT5 & 4.83 & 80.82\\
        CodeReviewer & 5.32 & 82.61\\
        \oapp & \textbf{7.33} & 82.84\\
    \bottomrule
\end{tabular}
\end{table}


\subsection*{Results for RQ3 - Effect of the embedding alignment objective}

\Table{tab:results4} illustrates the impact of the embeddings alignment objective on the performance of our proposed model \oapp~compared to baseline models.
As shown in this table, omitting this objective results in a drop in the model's performance (\ie the BLEU score drops from $7.33$ to $6.68$) for \emph{comment generation}.
However, \oapp~still has better results compared to the baselines even without the embedding alignment objective.
That is, a simple feedback signal from the teacher model to the student model may be sufficient to improve the model's accuracy. However, incorporating more informative feedback signals may improve further the model's accuracy.
This illustrates the importance of both components as each one contributes to the results' improvement.
Since the revised code and the review share the same meaning, converging to close representations resulted in boosting the performance of \oapp~on code review tasks.

The CodeBLEU score dropped slightly on the \emph{code refinement} task. This confirms our conjecture of RQ2 that the observed improvement in the \emph{code refinement} came from the embedding alignment. 

These findings suggest that aligning the embeddings for the different inputs that share similar semantics might be beneficial to the improvement of the model's performance on the downstream task.

\begin{table}[!htbp]
  \centering
  \caption{Influence of the \textit{embeddings alignment objective} on the performance of \oapp~for the comment generation task}
  \label{tab:results4}
  \begin{tabular}{l c c}
  \toprule
    \textbf{Model} & \textbf{Comment generation} & \textbf{Code refinement}\\
    \midrule
    \rowcolor{black!10} \multicolumn{3}{c}{\textcolor{black}{\textbf{Pre-finetuning}}}\\
        \oapp & & 81.79\\
    \rowcolor{black!10} \multicolumn{3}{c}{\textcolor{black}{\textbf{Fine-tuning}}}\\
        T5 & 4.39 & 77.03\\
        CodeT5 & 4.83 & 80.82\\
        CodeReviewer & 5.32 & 82.61\\
        \oapp & \textbf{6.68} & 80.96\\
    \bottomrule
\end{tabular}
\end{table}


\subsection*{Results for RQ4 - Training time of \oapp}

To evaluate the performance of \oapp~in comparison to baseline models, we recorded the training time required for each model to complete a single epoch. We employed four NVIDIA RTX-A5000 GPUs, each equipped with 24 gigabytes of dedicated memory. Careful consideration was given to configuring batch sizes to fully utilize these resources. Our aim was to assess the temporal efficiency and, by extension, the computational performance of all four models when trained for a single epoch using identical resources. The empirical results are detailed in \Table{tab:results6}.

\oapp~exhibited a substantial training time, consuming $411$ minutes, and extending to $524$ minutes when the embeddings alignment component was included, indicating relatively prolonged training durations. In stark contrast, the other models demonstrated remarkable computational efficiency, as they share the same architecture (i.e., the T5 architecture). While our proposed architecture outperforms the baseline models in terms of accuracy, it is worth noting that it demands more computational resources. However, it is important to emphasize that this resource-intensive phase occurs only once during training, posing no significant challenges in subsequent real-world deployment and usage.

\begin{table}[!htbp]
  \centering
  \caption{Performance of \oapp, in terms of training time per 1 epoch, compared to baseline models}
  \label{tab:results6}
  \begin{tabular}{l c c}
    \toprule
    \textbf{Model} & \textbf{Time (minutes)}\\
    \midrule
        T5 & 176\\
        CodeT5 & 179\\
        CodeReviewer & 178\\
        \oapp & 411\\
        \oapp + embeddings & 524\\
    \bottomrule
\end{tabular}
\end{table}

In the next research questions, we examine further our proposed architecture.
Given the evident resource-intensive nature of \oapp, we omit the embedding alignment objective.
Through this deliberate simplification, we would be able to run more experiments and shed light on the inherent potential of our architectural design thereby enabling a more nuanced evaluation of its performance and applicability.


\subsection*{Results for RQ5 - Accuracy on comment generation for each programming language}
\Table{tab:results2} shows the results of the performance of \oapp~model, compared to \emph{Code reviewer}, on the \emph{comment generation} task for each programming language. We perform the comparison with the baseline with the best accuracy, \ie, \emph{CodeReviewer}.

\oapp~outperforms \emph{CodeReviewer} on \emph{comment generation} for all the programming languages.
However, this advantage is variable from one programming language to another.
This depends on the quality and the amount of data for each programming language.
Also, software issues and best practices are different from one programming language to another.

We notice that \oapp~has a much better BLEU score for \emph{PHP}, and \emph{GO}.
However, the BLEU score is lower for \emph{C} and \emph{Java}.
This might depend on the usage and best practices for each programming language.
While some programming languages (\eg Java, C) have several best practices, coding standards, conventional issues, etc., developers may be more tolerant and lenient with other languages (\eg PHP).
For instance, the Java community has a strong focus on code quality and best practices. There are many established coding standards and guidelines, such as the Java Code Conventions, which encourage developers to write high-quality, and maintainable code.
This could lead to more strict code review processes to ensure that these standards are upheld.

Moreover, the performance variability could be related to the nature and characteristics of the programming language.
Some programming languages are compiled, which means that the program will not be functional if it has errors.
which could lead to more stringent code review processes.
Other languages are interpreted, which means they are more tolerant as the program will still be functional and errors may occur during run-time.

\begin{table*}[!htbp]
  \centering
  \caption{BLEU of \oapp, compared to baseline models, on comment generation per programming language}
  \label{tab:results2}
  \begin{tabular}{lccccccccc}
    \toprule
    \textbf{Language} & \textbf{PHP} & \textbf{Ruby} & \textbf{C\#} & \textbf{C} & \textbf{Java} & \textbf{Python} & \textbf{C++} & \textbf{Go} & \textbf{JavaScript}\\
    \midrule
    \textbf{CodeReviewer} & 8.10 & 5.49 & 5.92 & 5.16 & 4.03 & 4.36 & 5.71 & 6.05 & 5.24\\
    \bottomrule
    \textbf{\oapp} & 9.55 & 5.90 & 6.14 & 5.28 & 5.37 & 5.90 & 6.10 & 7.37 & 6.06\\
\end{tabular}
\end{table*}


\subsection*{Results for RQ6 - Impact of the teacher pre-finetuning}
In this research question, we investigated the impact of the teacher pre-finetuning phase on the performance of \oapp.
\Table{tab:results3} illustrates the influence of the pre-finetuning of the \emph{code refinement} model (\ie teacher) on the performance of \oapp~for the \emph{comment generation} task.

For \emph{comment generation}, there is an improvement of $0.46$ on the BLEU score when performing the pre-finetuning phase.
This demonstrates the importance of this phase as it enables the teacher model to acquire prior knowledge and have better performance on \emph{code refinement}.
Thus, the teacher would provide the student with more relevant feedback during the fine-tuning phase. Consequently, the student model would be able to generate more accurate comments that align with the reviewers' comments.

Still, both experiments produce better results for \emph{comment generation} compared to the literature.
This demonstrates the effectiveness of jointly addressing code review tasks using cross-task knowledge distillation.

These findings suggest that addressing several related tasks simultaneously using cross-task knowledge distillation might be beneficial even without having a very effective teacher model.
However, the pre-finetuning phase is important as it allows the teacher model to provide more informative feedback yielding better results for the student model.

\begin{table}[!htbp]
  \centering
  \caption{Impact of pre-finetuning the teacher (\ie code refinement model) on the performance of \oapp. }
  \label{tab:results3}
  \begin{tabular}{c c}
    \toprule
    \textbf{Comment generation} & \textbf{Code refinement}\\
    \midrule
    \rowcolor{black!10} \multicolumn{2}{c}{\textcolor{black}{\textbf{Without pre-finetuning}}}\\
        6.22 & 71.54\\
    \rowcolor{black!10} \multicolumn{2}{c}{\textcolor{black}{\textbf{With pre-finetuning}}}\\
        \textbf{6.68} & 82.84\\
    \bottomrule
\end{tabular}
\end{table}

\subsection{Threats to validity}
The evaluation results have demonstrated the effectiveness of our proposed architecture in addressing interconnected code review tasks. Nevertheless, there are certain threats that may limit the validity of the presented evaluation results.

A primary threat pertains to the nature of the data, specifically the reviews, which may contain noise and potentially non-English or misspelled words. We have mitigated this by employing codeT5, a state-of-the-art language model that utilizes Byte-Pair Encoding, a subword-based tokenization algorithm. This algorithm breaks unseen words into several frequently seen sub-words that can be effectively processed by the model.

A second concern revolves around the data imbalance, where some programming languages have fewer examples than others. This disparity may lead to varying performance across programming languages. However, the use of a large pre-trained language model allows us to circumvent this issue, as CodeT5 is a large language model that has already been trained on extensive code repositories. Consequently, fine-tuning this model for downstream tasks does not require a large volume of data. Moreover, prior research has shown that the use of multilingual training datasets can result in enhanced model performance compared to monolingual datasets, particularly for low-resource languages, in tasks such as neural machine translation and code translation \cite{chiang2021breaking, zhu2022multilingual}.

A final concern pertains to the selection of hyperparameters, which play a critical role in determining the model's performance. To ensure a fair comparison with \cite{li2022automating}, we conducted a grid search solely on the learning rate and batch size parameters. For other hyperparameters, we adopted the same settings as those used in \cite{wang2021codet5}. It is important to note that further improvements may be achievable through additional hyperparameter tuning.

\section{Related work}
\label{sec:literature}

Various techniques have been explored, in the literature, to support different tasks of code review, and these techniques can be classified into three categories: static analysis tools, similarity-based methods, and generative approaches.

In an effort to tackle the early stages of the code review process, a research direction has emerged that seeks to identify potential issues through the use of static analysis tools such as Checkstyle\cite{checkstyle}, PMD\cite{pmd}, and FindBugs\cite{hovemeyer2004finding, findBugs}. These tools, commonly referred to as linters, establish a set of rules that denote different types of issues related to security, code style, code quality, etc., and highlight parts of the code that are non-compliant with the defined rules.
Albeit the usefulness of such tools, the manual adaptation required to encompass most of the issues renders these tools rigid and less effective. Moreover, these tools need to be continuously calibrated with respect to issues and best practices that are variable over time and are affected by several factors such as software architecture, team composition, project characteristics, domain, and the like.
This inflexibility of static analysis tools diminishes their suitability for general adoption within software projects.

Other works employed similarity-based approaches to assist the code review process. These approaches assume that, in the context of code review, analogous situations may be encountered implying that similar issues may arise.
Consequently, historical knowledge can be exploited to hasten the code review process by resolving recurring issues.
Gupta et al. \cite{gupta2018intelligent} introduced DeepCodeReviewer (DCR), an LSTM-based model that is trained using positive and negative examples of (code, review) pairs. Given a new code snippet, a subset of candidate reviews is selected, from a predefined set of reviews, based on code similarity. Subsequently, DCR predicts a relevance score for each review with respect to the input code snippet and suggests reviews exhibiting high relevance scores.
Another work, proposed in \cite{siow2020core}, introduced a more sophisticated approach based on multi-level embedding to learn the relevancy between code and reviews. This approach uses word-level and character-level embeddings to achieve a better representation of the semantics provided by code and reviews.
These approaches have poor performance for unique code snippets and recommend irrelevant reviews in most cases. 

The last category, \ie generative approaches, attempted to use generative deep learning techniques to recommend reviews or code edits.
Tufano et al. \cite{tufan2021towards} proposed an approach that partially automates code corrections before and during code review. This approach is composed of two main components: a 1-encoder transformer that recommends additional changes to the developer before submitting her code and a 2-encoder transformer that suggests the necessary code edits to satisfy the reviewer comments \cite{tufan2021towards}.
The authors improved their approach, in \cite{tufano2022using}, using T5, a pre-trained text-to-text transfer transformer \cite{raffel2019exploring}. To provide the model with prior knowledge of the downstream tasks, the authors pre-trained the T5 model on Java and technical English datasets.
Li et al. \cite{li2022automating} pre-trained CodeT5 on four tasks, tailored specifically for the code review scenario, using a large-scale multilingual dataset of code reviews to better understand code differences and reviews. Then, the output model was fine-tuned on three downstream tasks: quality estimation (\ie accept/reject a pull request), review generation (\ie generate review comment), and code refinement (\ie recommend code edits to satisfy the reviewer comment).

Despite their tight relationship, these approaches addressed the different tasks of code review separately. In contrast, our approach employs code edits as feedback for the code review generation process. To achieve this, we rely on knowledge distillation introduced in \cite{hinton2015distilling}, where a model endowed with a greater number of parameters is employed to assist a smaller model in learning to produce decisions of comparable quality to the larger model, on the same task. More recently, cross-task distillation has been used in computer vision \cite{8890866, li2022prototypeguided, Ye_2020_CVPR}, allowing models to leverage knowledge gained from one task to improve performance on another related task. 
Yang et al. \cite{yang2022cross} proposed a cross-task knowledge distillation framework designed for multi-task recommendation. The framework seeks to utilize the prediction results of one task as supervised signals to instruct another task. The paper focuses on multi-task learning for predicting different user feedback (e.g. click, like, purchase, lookthrough).


\section{Conclusion}
\label{sec:conclusion}

We introduce \oapp, a novel deep learning architecture that leverages cross-task knowledge distillation to simultaneously tackle two interconnected tasks. This architecture comprises two models that undergo joint fine-tuning, guided by the aim of accomplishing both tasks. We have employed this framework to enhance the performance of deep learning models in automating code review tasks. The guidance within this architecture is implemented through two strategies: a feedback-based learning objective and an embedding alignment objective.

We evaluated \oapp~, with a dataset of code reviews, and compared it to state-of-the-art approaches. The results clearly demonstrate that \oapp~outperforms baseline models in both code review tasks: \emph{comment generation} and \emph{code refinement}. 
Furthermore, we investigated the impact of the embedding alignment objective and teacher pre-finetuning on \oapp's performance. Our findings illustrate that embedding alignment leads to further performance enhancements, as the code edits and reviews share similar meanings. Additionally, providing the teacher model with prior knowledge results in a more efficient feedback mechanism. These outcomes suggest that concurrently addressing multiple tasks can be advantageous, enabling the sharing of knowledge and facilitating the development of collaborative models that not only learn from ground truth but also incorporate feedback from other activities.

In our future work, we intend to design an alternative architecture with bidirectional feedback, wherein both models collaborate to achieve superior performance.
 Additionally, we plan to apply this generic framework to address other tasks with correlated activities. Finally, we aim to integrate our model as a bot within IDEs, allowing us to conduct user studies to assess its practical impact and usefulness from a developer's perspective.

\bibliographystyle{ACM-Reference-Format}
\bibliography{references}
\end{document}